\documentclass{aastex62}

\usepackage{epstopdf}
\usepackage{rotating}
\usepackage{subfigure}
\usepackage{graphicx}
\usepackage{float}

\newcommand{\upp}

\begin{document}
\title{Searching for Gravitationally Lensed Gamma-ray Bursts with Their Afterglows}

\correspondingauthor{He Gao}
\email{gaohe@bnu.edu.cn}

\author{Shengnan Chen}
\affil{Department of Astronomy ,
Beijing Normal University, Beijing 100875, China;}

\author{Xudong Wen}
\affil{School of Science ,
Wuhan University of Technology, Wuhan 430070, China;}

\author{He Gao}
\affiliation{Department of Astronomy ,
Beijing Normal University, Beijing 100875, China;}

\author{Kai Liao}
\affil{School of Science ,
Wuhan University of Technology, Wuhan 430070, China;}

\author{Liangduan Liu}
\affiliation{Department of Astronomy ,
Beijing Normal University, Beijing 100875, China;}

\author{Litao Zhao}
\affiliation{Department of Mathematics and Physics, Hebei GEO University, Shijiazhuang 050016, China;}

\author{Zhengxiang Li}
\affiliation{Department of Astronomy ,
Beijing Normal University, Beijing 100875, China;}

\author{Marek Biesiada}
\affiliation{Department of Astronomy ,
Beijing Normal University, Beijing 100875, China;}
\affiliation{National Centre for Nuclear Research ,
Pasteura 7, 02-093 Warsaw, Poland;}

\author{Aleksandra Pi\'orkowska-Kurpas}
\affiliation{Institute of Physics, University of Silesia, 75
Pu{\l}ku Piechoty 1, 41-500 Chorz{\'o}w, Poland}

\author{Shuo Xiao}
\affiliation{Institute of High Energy Physics (IHEP), Chinese Academy of Sciences (CAS), Beijing 100049, China}

\author{Shaolin Xiong}
\affiliation{Institute of High Energy Physics (IHEP), Chinese Academy of Sciences (CAS), Beijing 100049, China}

\begin{abstract}
Gamma-ray bursts (GRBs) at high redshifts are expected to be gravitationally lensed by objects of different mass scales. Besides a single recent claim, no lensed GRB has been detected so far by using the gamma-ray data only. In this paper, we suggest that the multi-band afterglow data might be an efficient way to search for lensed GRB events. Using the standard afterglow model we calculate the characteristics of the lensed afterglow lightcurves under the assumption of two popular analytic lens models: point mass and Singular Isothermal Sphere (SIS) model. In particular, when different lensed images cannot be resolved, their signals would be superimposed together with a given time delay. In this case, the X-ray afterglows are likely to contain several X-ray flares of similar width in linear scale and similar spectrum, and the optical afterglow lightcurve will show rebrightening signatures. Since the lightcurves from the image arriving later would be compressed and deformed in the logarithmic time scale, the larger time delay (i.e. the larger mass of the lens), the easier is to identify the lensing effect. We analyzed the archival data of optical afterglows and found one potential candidate of the lensed GRB (130831A) with time delay $\sim$500 s, however, observations of this event in gamma-ray and X-ray band seem not to support the lensing hypothesis. In the future, with the cooperation of the all-sky monitoring gamma-ray detectors and multi-band sky survey projects, our method proposed in this paper would be more efficient in searching for strongly lensed GRBs.
\end{abstract}


\section{Introduction} \label{sec:intro}

As one of the most violent explosions in the Universe, GRBs and their afterglows are bright enough to be detected in high-redshift range up to at least $z\sim10$ \citep{tanvir09,salvaterra09}. In view of their high redshift, the gravitational lensing effect on GRBs has long been discussed ever since the cosmological origin for GRBs was first proposed \citep{paczynski86}. With detailed study on the probability distributions of time delay in gravitational lensing by point masses and isolated galaxies (modeled as singular isothermal spheres), \cite{mao92} predicted that the probability of multiple GRB images obtained by galactic lensing event could be between $0.05\%$ and $0.4\%$. And by assuming that the GRB luminosity function has no cosmological evolution and the lens is modeled as singular isothermal spheres, \cite{hurley19} estimated that the detection rate of both lensed GRB images is roughly 0.05 times per year.

Thanks to the successful operation of several dedicated detectors, e.g., the Burst And Transient Source Experiment (BATSE) on Compton Gamma Ray Observatory (CGRO) \citep{meegan92}, the Burst Alert Telescope (BAT) on the Neil Gehrels Swift Observatory \citep{gehrels04,barthelmy05} and the Gamma-Ray Burst Monitor (GBM) on the Fermi Observatory \citep{meegan09}, $\sim10^4$ GRBs have been detected. The searches for gravitational lensing of GRBs have been widely carried out, for instance, \cite{li14} searched $\sim$2100 GRBs observed by BATSE, \cite{hurley19} searched $\sim$2300 GRBs observed by Konus-Wind, and \cite{ahlgren20} searched $\sim$ 2700 GRBs observed by GBM. In addition, \cite{veres09} and \cite{davidson11} also failed to search for lensing events with two smaller samples of GRBs observed in the first years of Fermi-GBM.

Unfortunately, all searches for galaxy lensing events (with time delay in order of days) have yielded null results. Most recently, \cite{paynter21} claimed that they had found one statistically significant lensing event in the lightcurve of GRB 950830 (BATSE trigger 3770) with sub-second time delay. The inferred lens mass, although depending on the unknown lens redshift, falls into the mass range of intermediate-mass black holes (e.g. $\sim10^{4}-10^{5}M_{\odot}$). The mismatch between theoretical predictions concerning lensing rates and the data search results might be due to the uncertainty of the theoretical model, or it may also be due to the fact that current gamma-ray detectors do not yet have all-sky monitoring \citep{li14}. Nevertheless, it is also possible that some real lensing events are missed in the data search process, considering that the current searches are mainly based on the comparison of gamma-ray lightcurves and the coincidence degree of positions, but 1) the radiation time scale of GRBs in gamma-ray bands is relatively short, 2) the shape of lightcurve is greatly affected by the background, and 3) the positioning error of gamma-ray detectors is very large.

Here we propose that in addition to looking for lensed GRBs in the gamma-ray band, we can also search them with GRB's afterglow observations, taking into account that the afterglow's detectable timescale is much longer (up to years), their lightcurve shape is relatively simple, their physical model is much more clear, and they can be monitored by either satellites or ground based multi-band detectors in order to enhance the probability of capturing the lensing effect  \cite[][for a review]{zhang18}. For strongly gravitational lensing events with lens mass $M_{l}$, the typical angular separation between multiple images could be roughly estimated as $0.1''{({M_l}/{10^{10}}{M_ \odot })^{1/2}}{({D_l}{D_s}/{D_{ls}}/{\rm{Gpc}})^{ - 1/2}}$ \citep{meng16}, which could hardly be resolved by facilities that are commonly used to observe GRB afterglows, unless the lens galaxy is extremely massive (e.g. $M_l>10^{12}M_{\odot}$). For instance, the angular resolution of the Swift X-ray Telescope (XRT) is only $18''$ \citep{burrows05} and the resolution $1$ meter-class optical telescopes is  $\sim1''$ \citep{oguri10}. On the other hand, by studying the broad time delay probability distribution of the point lens, \cite{mao92} found that the peak of the time delay probability appeared at $50 \hspace{0.04cm} {\rm{s}}\;({M_l}/{10^6}{M_ \odot })$.  Unlike the short-term gamma-ray emission, the detectable timescale for afterglows is typically much longer than the time delay timescale. In this case, if the multiple images are unresolved, the signals of multiple images will be superimposed to disguise as one signal.

In this paper, we first use the standard GRB afterglow and gravitational lensing models to calculate the lightcurve of such superimposed signal in different bands. Based on the characteristics of the calculated lightcurves, we then searched the current GRB afterglow observation data in the optical band. One potential candidate, GRB 130831A is found, and the possible parameter space is investigated in order to check if it is indeed a lensed event. Future prospects are discussed in the final section. Throughout this paper, we assume the Planck cosmology \citep{planck15} as a fiducial model, with ${\Omega _{\rm{m}}} = 0.307$, ${\Omega _\Lambda } = 0.693$ and ${H_0} = 67$ ${\rm{km}}$ ${{\rm{s}}^{ - 1}}$ ${\rm{Mp}}{{\rm{c}}^{ - 1}}$.

\section{GRB afterglow Lightcurves with Unresolved Lensing Effect}

\subsection{GRB Afterglow Model}

Although the nature of GRB's progenitor and central engine as well as the detailed physics of gamma-ray emission are still rather uncertain, a generic synchrotron external shock model has been well established to describe the interaction between the relativistic GRB jet and the circumburst medium, thus to interpret the broad-band afterglow data \cite[see][for a review]{gao13}. The total effective kinetic energy of the jet and the medium can be expressed as
\begin{equation}
E_k = {\rm{ }}\left( {\Gamma  - 1} \right){M_{\rm ej}}{c^2} + \left( {{\Gamma ^2} - 1} \right){M_{\rm sw}}{c^2},
\end{equation}
where $\Gamma $ is the bulk Lorentz factor of the outflow and ${M_{\rm sw}}$ is the swept-up mass by the shock. The initial mass ejected from the central engine, ${M_{\rm ej}}$, is a fixed value determined by ${M_{\rm ej}} = ({E_0}(1 - \cos{\theta _0}))/({\Gamma _0}{c^2})$, where ${E_0}$, and ${\Gamma _0}$, ${\theta _0}$ are  the initial total kinetic energy, initial bulk Lorentz factor and the initial opening angle of the jet, respectively. Without considering the case of energy injection and energy loss due to shock emission, the differential form of energy conservation is given by \citep{huang20}
\begin{equation}
\frac{{d \hspace{0.03cm}\Gamma }}{{d{M_{\rm sw}}}} =  - \frac{{{\Gamma ^2} - 1}}{{{M_{\rm ej}} + 2\Gamma {M_{\rm sw}}}}.
\end{equation}
Applying the evolution expressions of ${M_{\rm sw}}$, $\theta$ and the radius $R$ of the blastwave \citep{huang20},
\begin{equation}
\frac{{d{M_{\rm sw}}}}{{dR}} = 2\pi {R^2}(1 - \cos\theta ){m_p}n,
\end{equation}
where $n$ is the number density of the circumburst medium,
\begin{equation}
\frac{{d\theta }}{{dt}} = \frac{{{c_s}\left(\Gamma  + \sqrt {{\Gamma ^2} - 1} \right)}}{R},
\end{equation}
and the sound speed ${{c_s}}$ can be calculated as
\begin{equation}
c_s^2 = \frac{\widehat \gamma (\widehat \gamma  - 1)(\Gamma  - 1)}{{1 + \widehat \gamma (\Gamma  - 1)}} \hspace{0.05 cm}{c^2},
\end{equation}
where $\widehat \gamma $ is the adiabatic index defined as $\widehat \gamma  = (4\Gamma  + 1)/(3\Gamma )$,
\begin{equation}
\frac{{dR}}{{dt}} = \sqrt {\frac{{{\Gamma ^2} - 1}}{\Gamma }} c\Gamma \left(\Gamma  + \sqrt {{\Gamma ^2} - 1} \right).
\end{equation}
One can then solve the dynamical evolution of the external shock numerically. During the shock propagation, electrons are accelerated in the shock, which will radiate synchrotron emission in the magnetic fields behind the shock that are believed to be generated in situ due to plasma instabilities \citep{medvedev99}.

The electron energy distribution accelerated by shock is usually considered as a power law distribution: ${N_{{\gamma _e}}}d{\gamma _e} \propto \nolinebreak {\gamma _e}^{ - p}d{\gamma _e}~(p>2)$. Assuming that a constant fraction ${\epsilon _e}$ of the shock energy goes into the electrons, combined with electric neutral conditions and shock jump conditions, the minimum injected electron Lorentz factor can be estimated as \citep{sari98}
\begin{equation}
\gamma_{\rm e,m} = {\epsilon _e} \left(\frac{{p - 2}}{{p - 1}}\right) \frac{{{m_p}}}{{{m_e}}}(\Gamma  - 1).
\end{equation}
Assuming a constant fraction $\epsilon_B$ of the shock energy density goes into the magnetic field, the magnetic field strength $B'$ in the comoving ejecta frame can be estimated as

\begin{equation}
B' \approx \sqrt {32\pi {c^2} {\epsilon _B}{\Gamma ^2}{m_p}n}.
\end{equation}
Synchrotron radiation power in the co-moving frame could then be calculated using the following formula  \citep{rybicki79}
\begin{equation}
\label{eq:pnup}
P'_{\nu'} = \sqrt{3}\hspace{0.1cm} \frac{q_e^3 B'}{m_{\rm e} c^2}
	    \int_{\gamma_{\rm e,m}}^{\gamma_{\rm e,M}}
	    \left( \frac{dN_{\rm e}'}{d\gamma_{\rm e}} \right)
	    F\left(\frac{\nu '}{\nu_{\rm cr}'} \right) d\gamma_{\rm e},
\end{equation}
where $q_e$ is electron charge, $\gamma_{\rm e,M}$ is the the maximum electron Lorentz factor that could be estimated by balancing the acceleration time scale and the dynamical time scale,
$\nu_{\rm cr}' = 3 \gamma_{\rm e}^2 q_e B' / (4 \pi m_{\rm e} c)$ is the
characteristic frequency of an electron with Lorentz factor $\gamma_e$, and
\begin{equation}
\label{fx23}
F(x) = x \int_{x}^{+ \infty} K_{5/3}(k) dk,
\end{equation}
where  $K_{5/3}(k)$ is the modified Bessel function of order $5/3$.


\subsection{Gravitational Lensing Models}

For the purpose of this work, we adopt two commonly used analytic lens models: point mass model and SIS model. Their choice is dictated by two realistic scenarios: lensing by an isolated mass -- presumably massive (with $M_l<10^{7} M_{\odot}$) black hole (BH) or an intervening galaxy ($M_l \sim 10^{9} - 10^{11} M_{\odot}$).
We will not discuss the details of these scenarios here, they are beyond the scope of this study.

In the case of axisymmetric point lenses, the typical separation between different images is set by the Einstein radius
\begin{equation}
{\theta _E} = \sqrt {\frac{{4G{M_l}}}{{{c^2}}}\frac{{{D_{ls}}}}{{{D_l}{D_s}}}} ,
\end{equation}
where $M_l$ is the mass of the lens, ${D_l}$ is the angular diameter distance to the lens at redshift ${z_l}$, ${{D_s}}$ is the angular diameter distance to the source at redshift ${z_s}$, ${D_{ls}}$ is the angular diameter distance between the lens and the source. The dimensionless lens equation \citep{Falco} and its corresponding solutions are

\begin{equation}
y = x - \frac{1}{x}\rightarrow \begin{array}{*{20}{c}}
{\left\{ {\begin{array}{*{20}{c}}
{{x_ + } = \frac{{y + \sqrt {{y^2} + 4} }}{2}}\\
{{x_ - } = \frac{{y - \sqrt {{y^2} + 4} }}{2}}
\end{array}} \right.}
\end{array}
\end{equation}
where $y = \beta /{\theta _E}$, $\beta$ denoting the angle between directions to the lens and to the source,  $x = \theta /{\theta _E}$, where $\theta $ is the angular position of the image actually seen by the observer. Angular positions of the source and its images obey the lens equation: $\beta  = \theta  - \alpha $, where $\alpha  = {{{D_{ls}}}\hat \alpha}/{{{D_s}}} $ is the reduced deflection angle. In the case of point mass lenses, two images could be produced at angular positions ${\theta _ + } = {x_ + }{\theta _E}$ and ${\theta _ - } = {x_ - }{\theta _E}$, respectively. The inverse of the determinant of the Jacobi matrix $\frac{\partial \beta}{\partial \theta}$ defines their magnifications,
\begin{equation}
{\mu _ \pm } = \frac{1}{1 - {\left(\frac{1}{{{x_ \pm }}}\right)^4}}.
\end{equation}
Time delay between images produced by a point mass lens is \citep{Falco}:
\begin{equation}
\Delta t = \frac{{4G{M_l}(1 + {z_l})}}{{{c^3}}} \left( \frac{{y\sqrt {{y^2} + 4} }}{2} + \ln \frac{{\sqrt {{y^2} + 4}  + y}}{{\sqrt {{y^2} + 4}  - y}} \right).{\rm{ }}
\end{equation}
Point mass lenses (also called Schwarzschild lenses) are representative to lensing by stars (microlensing) and BHs with stellar/intermediate mass.

It is well established that early-type galaxies act as lenses in the majority of strongly gravitational lens systems detected. Even though their formation and evolution are still not fully understood in details, a singular isothermal sphere model (SIS) can reasonably characterize the mass distribution of massive elliptical galaxies within the effective radius
\citep{Treu04,Treu06,Cao16,Liu20}. In the case of axisymmetric SIS model \citep{nara96,bern99}, its three-dimensional density profile could be described as $\rho (r) = {{\sigma _v^2}}/{{2\pi G{r^2}}}$, where $r$ is the distance from the sphere center and ${\sigma _v}$ is the velocity dispersion of the lens. By projecting the three-dimensional density along the line of sight, we obtain the corresponding surface density
\begin{equation}
\Sigma(\xi ) = 2 \frac{{\sigma _v^2}}{{2\pi G}}\int_0^\infty  {\frac{{dz}}{{{\xi ^2} + {z^2}}}}  = \frac{{\sigma _v^2}}{{2G\xi }},
\end{equation}
where $\xi  = \theta {D_l}$ is the impact parameter on the image plane. In this case, the dimensionless lens equation \citep{Falco} could be written as
\begin{equation}
y = x - \frac{x}{{|x|}},
\end{equation}
where $x = {\xi }/{{{\xi _0}}}$, ${\xi _0} = 4\pi {({{{\sigma _v}}}/{c})^2}{{{D_l}{D_{ls}}}}/{{{D_s}}}$ is the length scale on the lens plane, $y = {\eta }/{{{\eta _0}}}$, where $\eta {\rm{ = }}\beta {D_s}$ and ${\eta _0} = {{{\xi _0}{D_s}}}/{{{D_l}}}$ is the length scale related to ${{\xi _0}}$ on the source plane. There are two cases for the solution of the lens equation: when $y < 1$, the solutions are
\begin{equation}
x _ {\pm}  = y \pm 1.
\end{equation}
The corresponding angular positions of the images are  ${\theta _ \pm } = \beta  \pm {\theta _E}$, where the Einstein radius ${\theta _E}$ could be written as
\begin{equation}
{\theta _E} = \sqrt {\frac{{4GM(\theta _E)}}{{{c^2}}}\frac{{{D_{ls}}}}{{{D_l}{D_s}}}} =  4 \pi \frac{\sigma_{v}^2}{c^2}
\frac{D_{ls}}{D_s},
\end{equation}
where $M(\theta _E)$ is the mass within the Einstein radius. In this case, the magnifications of the images are
\begin{equation}
\mu _ {\pm}  = 1 \pm \frac{1}{y}.
\end{equation}
On the other hand, when $y > 1$, the lens equation only has one solution: $x=y+1$. Its corresponding magnification is $\mu  = {{|x|}}/({{|x| - 1}})$.

For SIS model, the time delay between different images reads:
\begin{equation}
\Delta t = \frac{32 \pi^2}{c} \left(\frac{{{\sigma _v}}}{c}\right)^4  \frac{{{D_l}{D_{ls}}}}{{{D_s}}}(1 + {z_l})y.
\end{equation}

\begin{figure}[hbt]
         \figurenum{1}
        \graphicspath{{Fit/}}
	\centering
	\figurenum{1}
         \includegraphics[width=8.5cm,height=6.9cm]{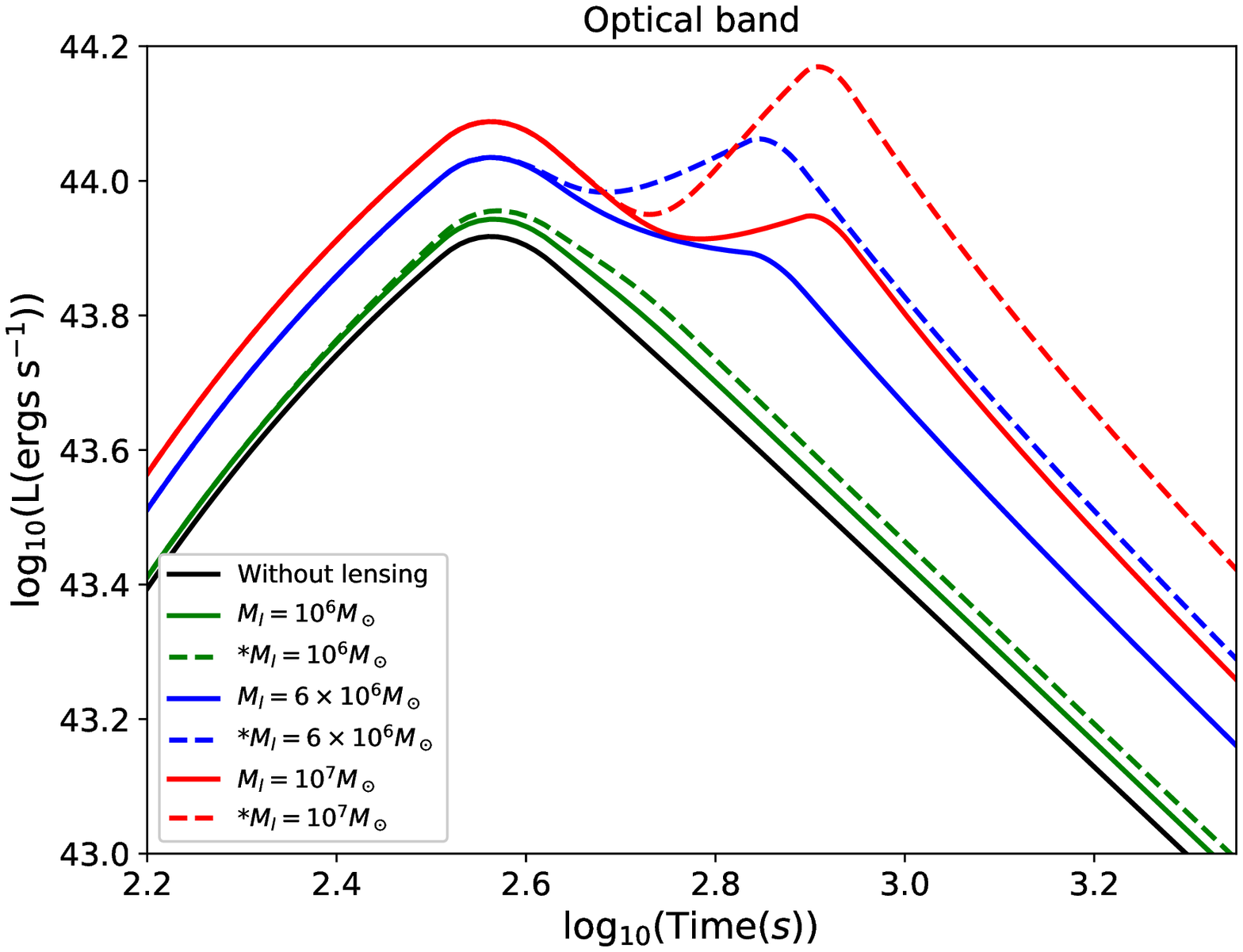}
         \includegraphics[width=8.5cm,height=6.9cm]{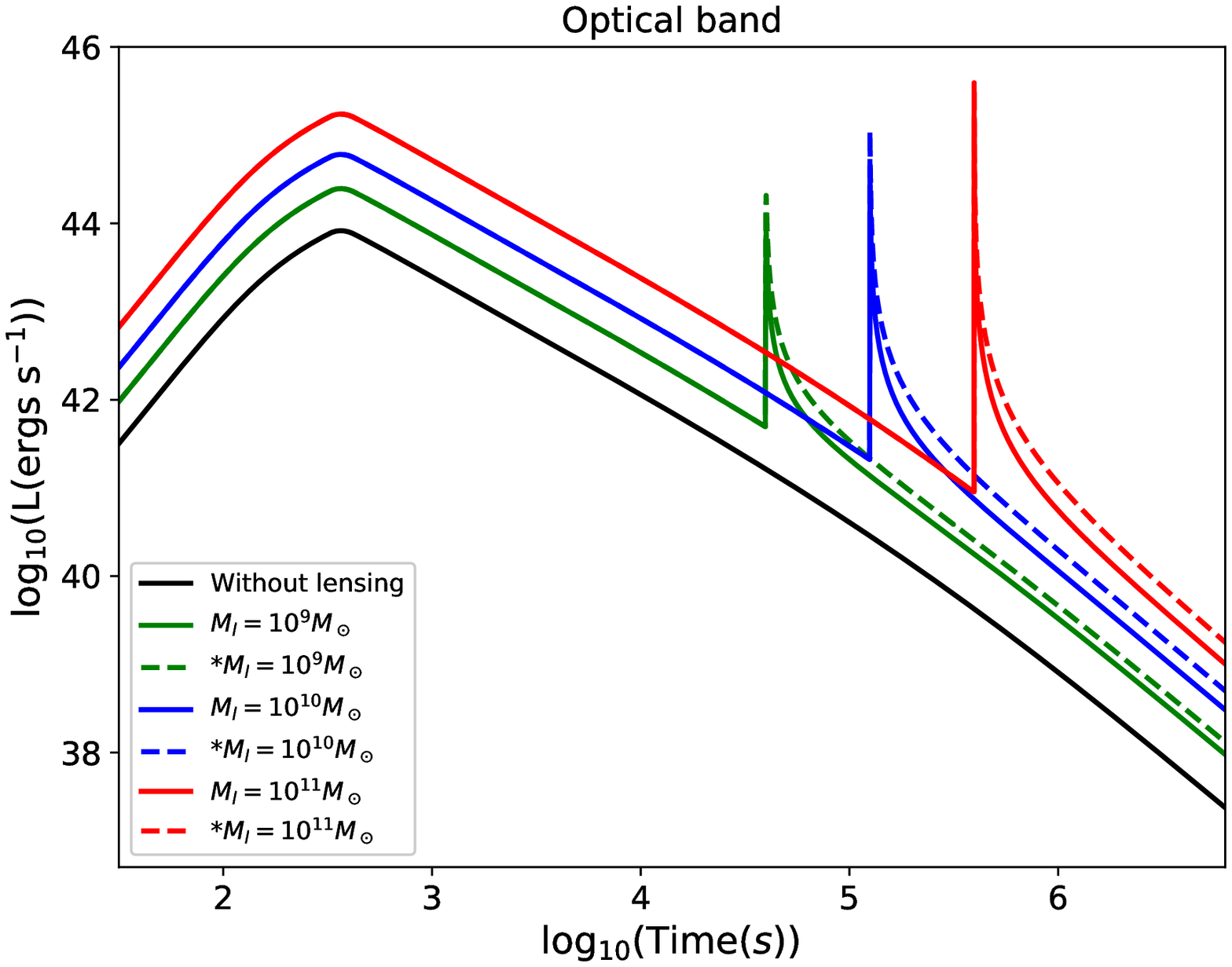}
         \caption{The lensed afterglow lightcurves in optical band when the lens mass is less than ${10^7}{M_ \odot }$ (the left panel) and larger than ${10^7}{M_ \odot }$ (the right panel) respectively. Different from the right panel where $t_{\rm delay}$ is larger than $t_{\rm peak}$, the green, blue and red lines in the left panel represent the cases where $t_{\rm delay}$ is smaller, equal or larger than $t_{\rm peak}$ respectively. All dotted lines marked with $*$ in legend are the result of additional magnification (here we manually set the additional magnification factor as $\mu_{add} =2.5$) of the second image by microlensing effect. The typical values for the model parameters: $E_{0}=10^{52}\ \rm{ergs}$,  $\Gamma_{0}=100$, $n =1\ \rm{cm}^{-3}$, $\theta_{0}=0.1$, $\epsilon_{e}=0.1$, $\epsilon_{B}=10^{-4}$, $p=2.5$, ${z_l} = 1$, ${z_s} = 2$, ${\beta _{{\rm{point}}}} = 1.8{\theta _{\rm{E}}}({M_l} = {10^6}{M_ \odot })$, ${\beta _{{\rm{SIS}}}} = 0.5\frac{{{\xi _0}}}{{{{\rm{D}}_{\rm{l}}}}}({M_l} = {10^{9}}{M_ \odot })$.}
	     \label{Superpositionone}
\end{figure}

\subsection{Superposition of Lensed Images}

When the lensed images are unresolved, our detected afterglow signal would be the superposition of signals from multiple images. In this case, the shape of afterglow lightcurve  depends on the number of images, their magnifications and time delay between different images.

In the optical band, the afterglow lightcurve is mainly contributed by the external shock emission. With the standard afterglow and gravitational lensing models as described above,  we calculated some optical lightcurves of such superimposed signal in various scenarios. There are several free parameters involved in our calculations, which can be divided into two categories. The first category is associated with the external shock, including initial kinetic energy of GRB outflow $E_{0}$, the initial bulk Lorentz factor of GRB outflow $\Gamma_{0}$, the initial opening angle of the jet $\theta_{0}$, the equipartition parameters for the magnetic field and electrons: $\epsilon_{B}$ and $\epsilon_{e}$, and the electron distribution index $p$. The second category is related to the lensing model, including the lens redshift ${z_l}$, the source redshift ${z_s}$, the angular position of the source $\beta $ and the lens mass ${M_l}$ in the point mass model and the SIS model. In SIS model mass within the Einstein radius is determined in the velocity dispersion and distance terms: $M_l = \frac{4 \pi^2 c^2}{G} \left(\frac{\sigma_v}{c}\right)^4 \frac{D_{ls}D_l}{D_s}$.

For the purpose of this work, we first fix the parameters in the first category at their typical values, e.g. we set $E_{0}=10^{52}\ \rm{ergs}$,  $\Gamma_{0}=100$, $n =1\ \rm{cm}^{-3}$, $\theta_{0}=0.1$, $\epsilon_{e}=0.1$, $\epsilon_{B}=10^{-4}$, $p=2.5$. With these settings, the afterglow lightcurves for each image is rather simple: corresponding to their own starting point, the lightcurve  rises as ${t^{2}}$, peaks around $t_{\rm peak}\sim360$ s and then declines as $t^{-1.2}$.

\begin{figure}[hbt]
         \figurenum{2}
        \graphicspath{{Fit/}}
	\figurenum{2}
       \includegraphics[width=6cm,height=4.8cm]{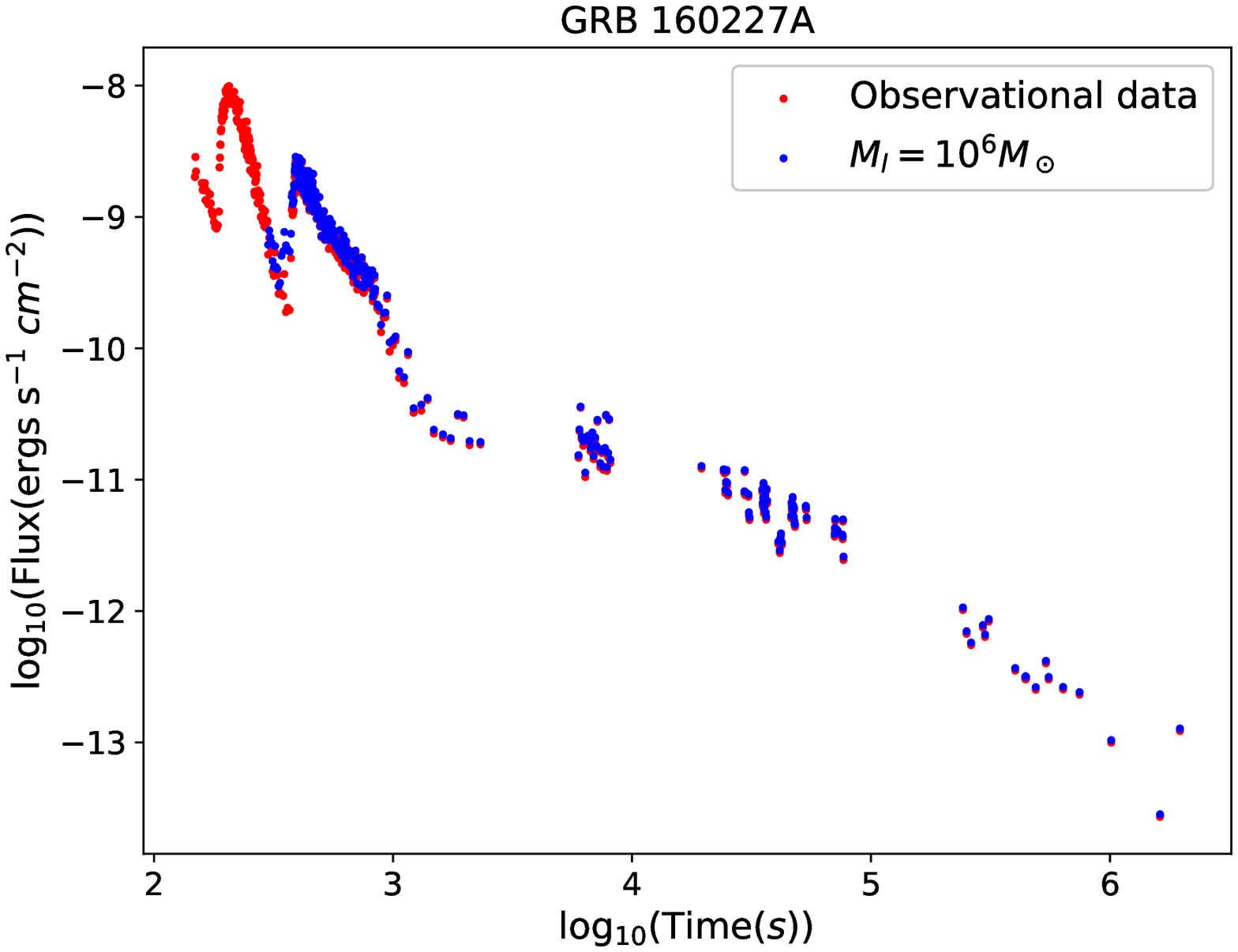}
       \includegraphics[width=6cm,height=4.8cm]{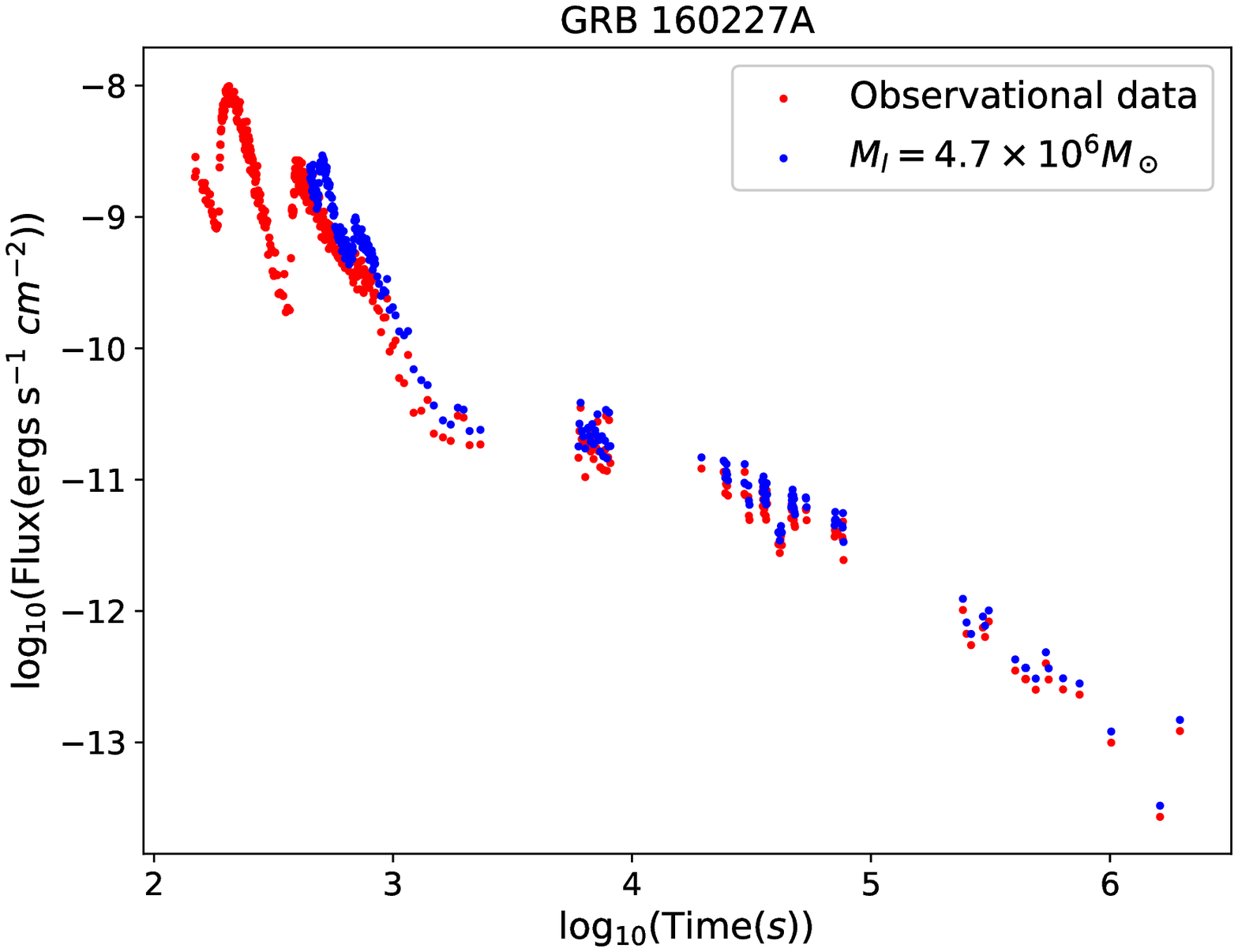}
       \includegraphics[width=6cm,height=4.8cm]{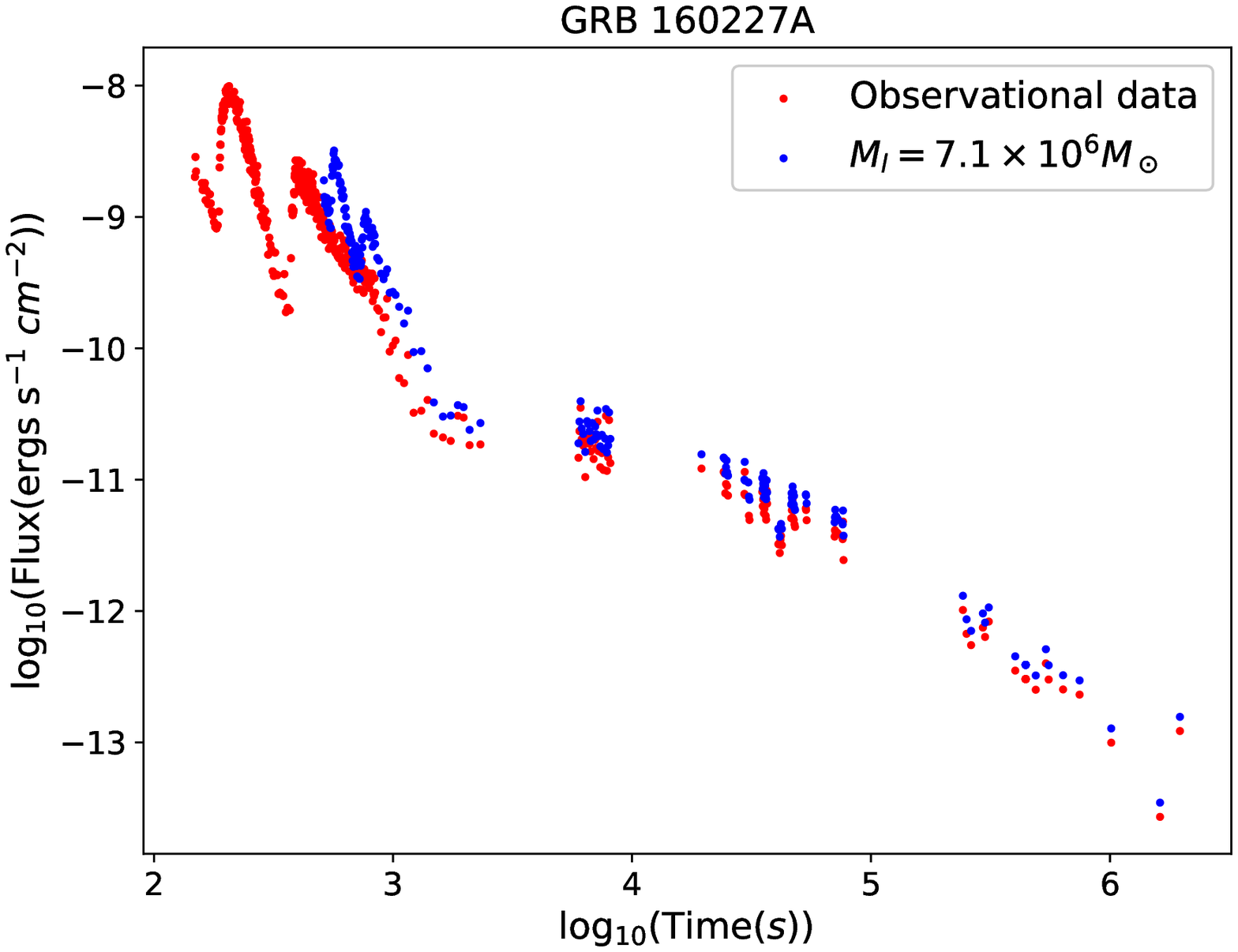}
       \includegraphics[width=6cm,height=4.8cm]{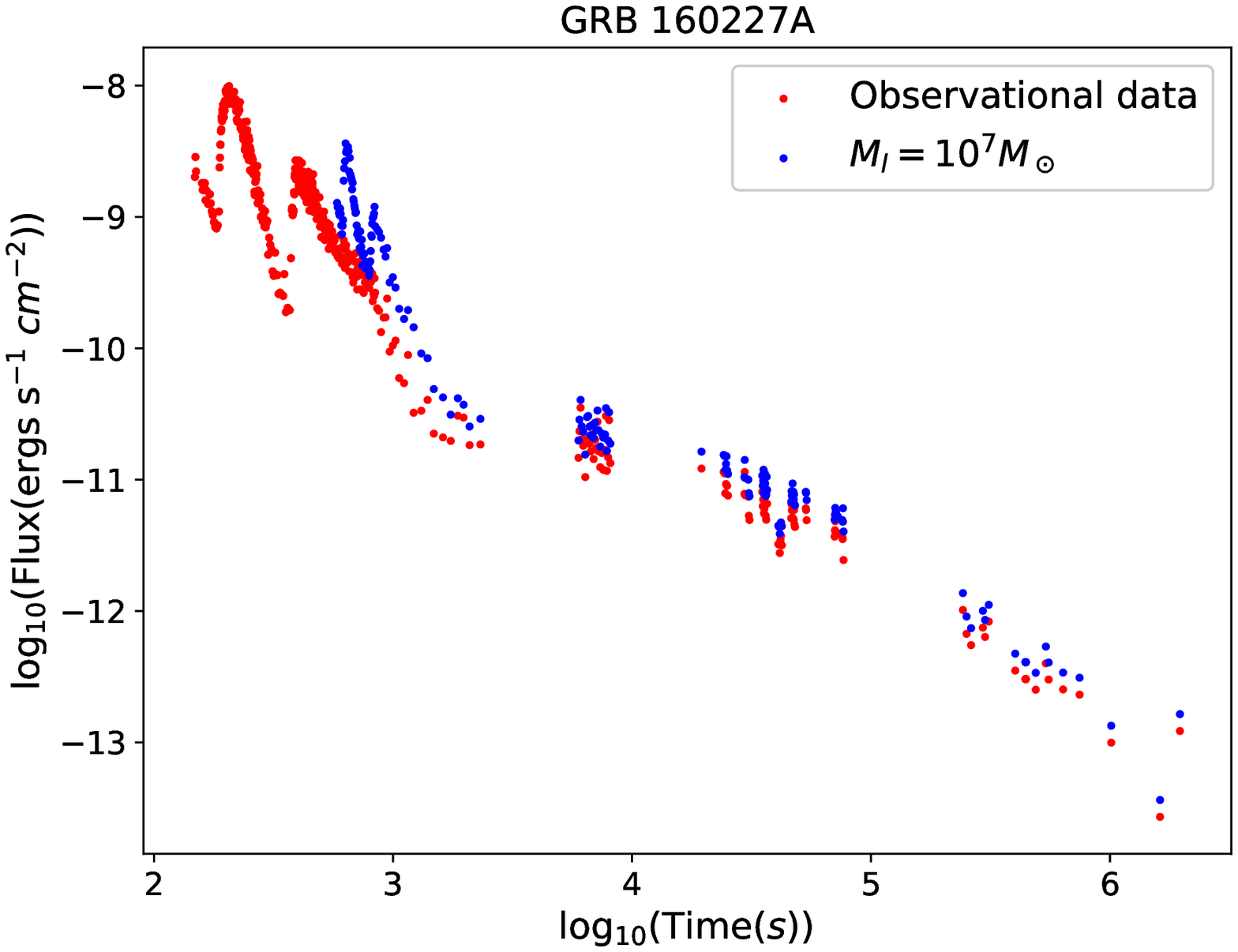}
       \includegraphics[width=6cm,height=4.8cm]{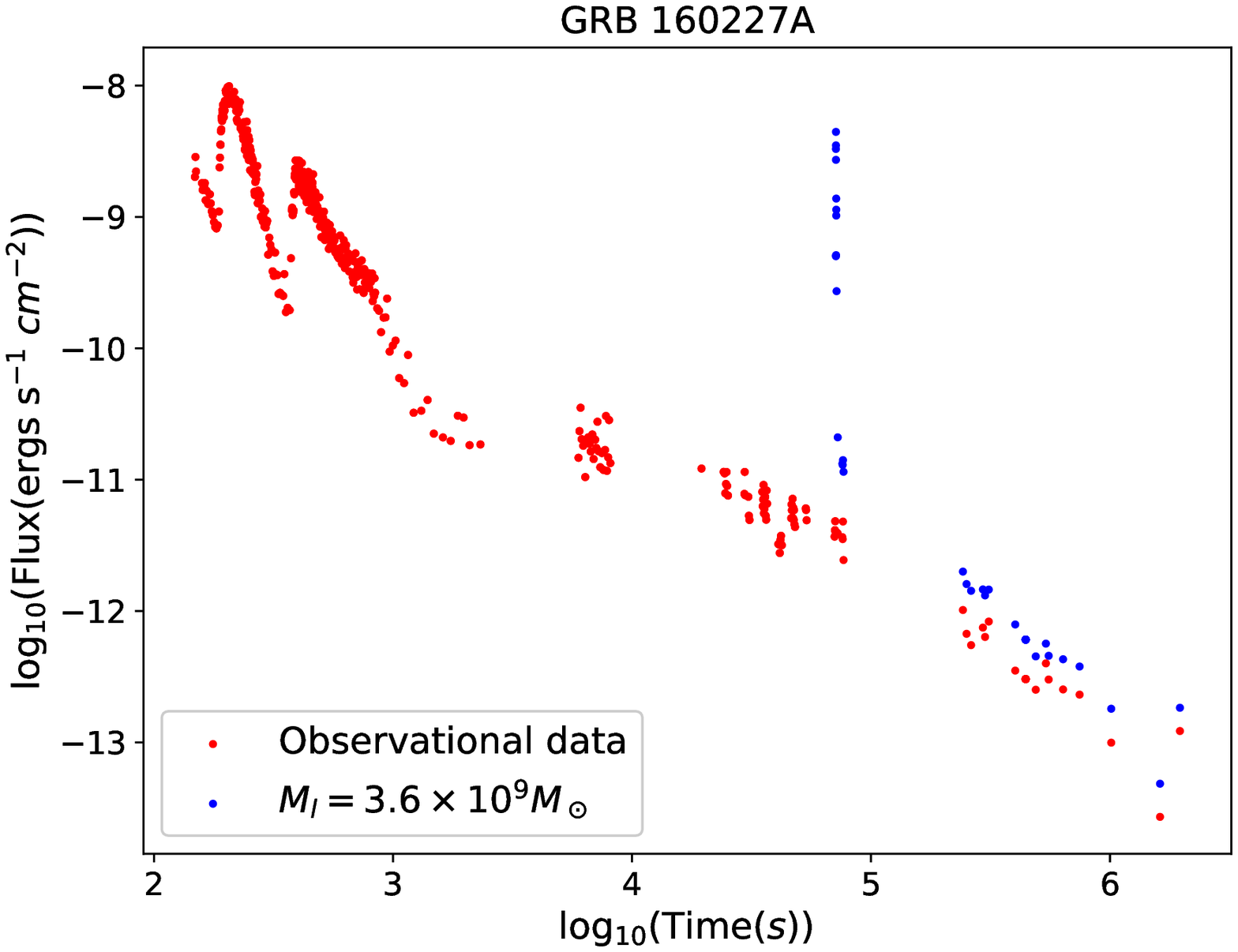}
       \includegraphics[width=6cm,height=4.8cm]{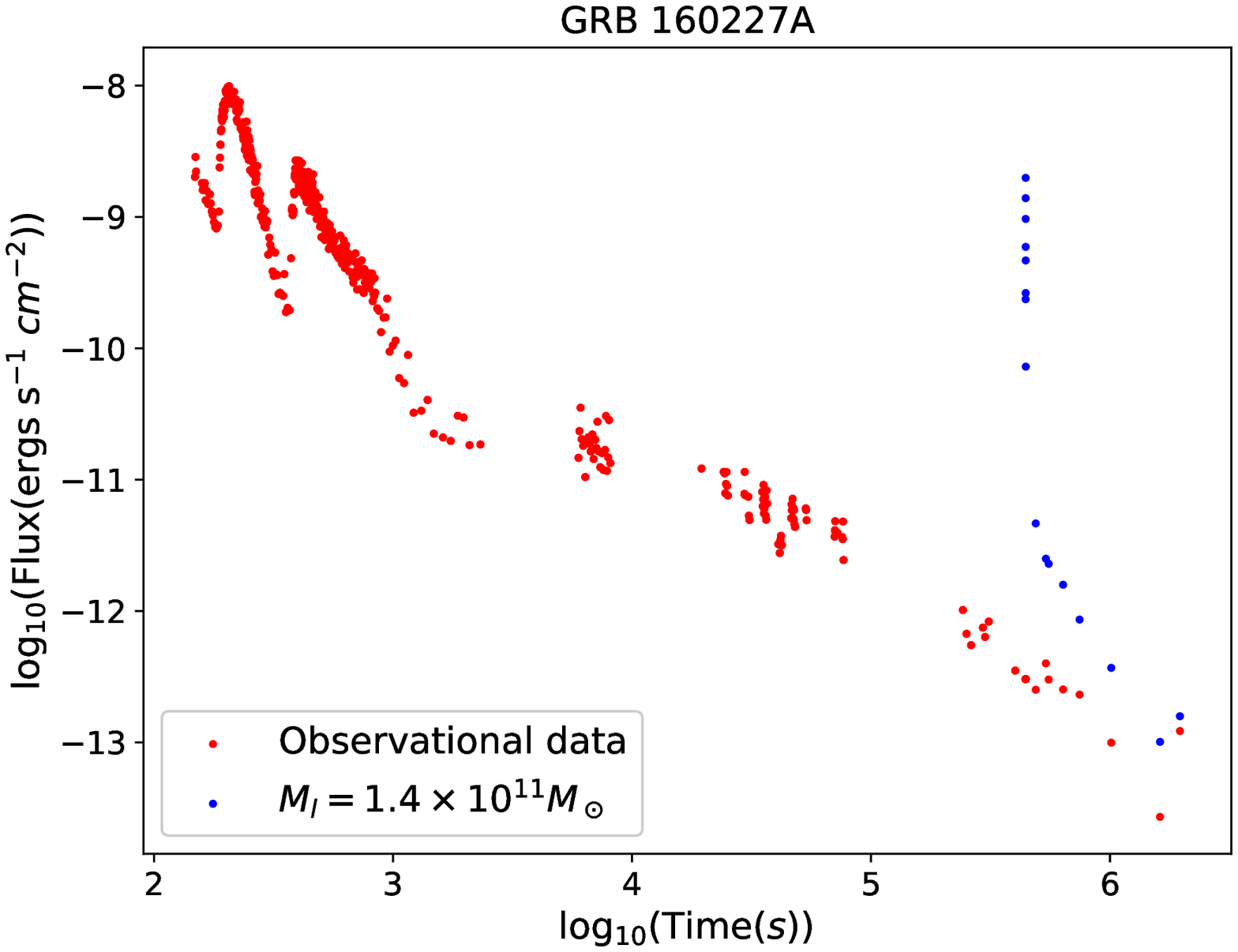}
       \caption{Simulation of gravitational lensing effect in X-ray afterglow of GRB 160227A. The observational data are marked with red dots. The blue dots are the result of the superposition from different lensed images after the second lensed image arrives. Here we set the flux from the first arrived image equaling to the observational data.}
	   \label{Superpositiontwo}
\end{figure}

By adjusting the values of parameters in the second category, we produced  various scenarios with $t_{\rm delay}$ (delay time between different lensed images) being smaller, equal or larger than $t_{\rm peak}$. As shown in Figure \ref{Superpositionone}, when we take the initial time of the first image as the starting time, the lightcurves from later arriving image would be compressed and deformed in the logarithmic time scale, which would behave as a rebrightening signature. When the lens mass is relatively small (e.g. $M_l<10^{7}{M_ \odot }$), $t_{\rm delay}$ is close to $t_{\rm peak}$. In this case, the rebrightening signature is not significant, unless the second image suffers additional magnification effects (such as microlensing). However, when the lens mass is large enough (e.g. $M_l>10^{7}{M_ \odot }$), $t_{\rm delay}$ is much larger than $t_{\rm peak}$ and in this case, the rebrightening signature will have an extremely sharp rise, which would be very easy to identify.

Unlike the optical band, X-ray afterglow is a superposition of the conventional external shock component and a radiation component that is related to the late central engine activity, manifested through flares and extended shallow plateaus \citep{zhang06}. The lensing effect for external shock component should be the same with optical band (as shown in Figure \ref{Superpositionone}). For the latter component, here we adopt the real observational data for GRB 160227A, which contains both X-ray flare and plateau, to simulate a pseudo lightcurve to show the lensing effects for flares and plateaus. As shown in Figure \ref{Superpositiontwo}, when $t_{\rm delay}$ is comparable to the timescale of early X-ray flares, the signal from the later arrived image will be masked as new flare components, superimposed with the flare of the first image or separated into multiple independent flares. In this case, the plateau signal does not change obviously, but the overall flux increases a little higher. When $t_{\rm delay}$ is large enough, the early X-ray signal would be compressed into a narrow jump pulse with extremely sharp rising and declining, which would be easy to identify once appears.

In the radio band, the afterglow lightcurve would peak at relatively late timescale. When the lens mass is small, $t_{\rm delay}$ would be much smaller than $t_{\rm peak}$, no obvious lensing feature would show up since signals of both images are in the rising stage and are relatively weak. On the other hand, when the lens mass is large enough to make $t_{\rm delay}$ close to or larger than $t_{\rm peak}$, the angular separation between multiple images could be larger than the angular resolution of radio telescopes. For example, the Very Large Array (VLA) with the minimum resolution of $0.04''$ \footnote{\url{https://public.nrao.edu/telescopes/vla/}}could in principle distinguish lensed GRB images when the lens mass is larger than ${10^9}{M_ \odot }$ (with $t_{\rm delay}>10^4$ s), so that the afterglow lightcurve from different images could be decomposed instead of superimposing. In this case, one can search directly for strongly gravitational lensed GRB in radio imaging. For these reasons, we did not show the results for radio band here.

\section{Archive data searching for lensed GRB candidates in optical band}

\begin{figure}[hbt]
         \figurenum{3}
        \graphicspath{{Fit/}}
    \centering
	\figurenum{3}
    	\includegraphics[width=8.5cm,height=7cm]{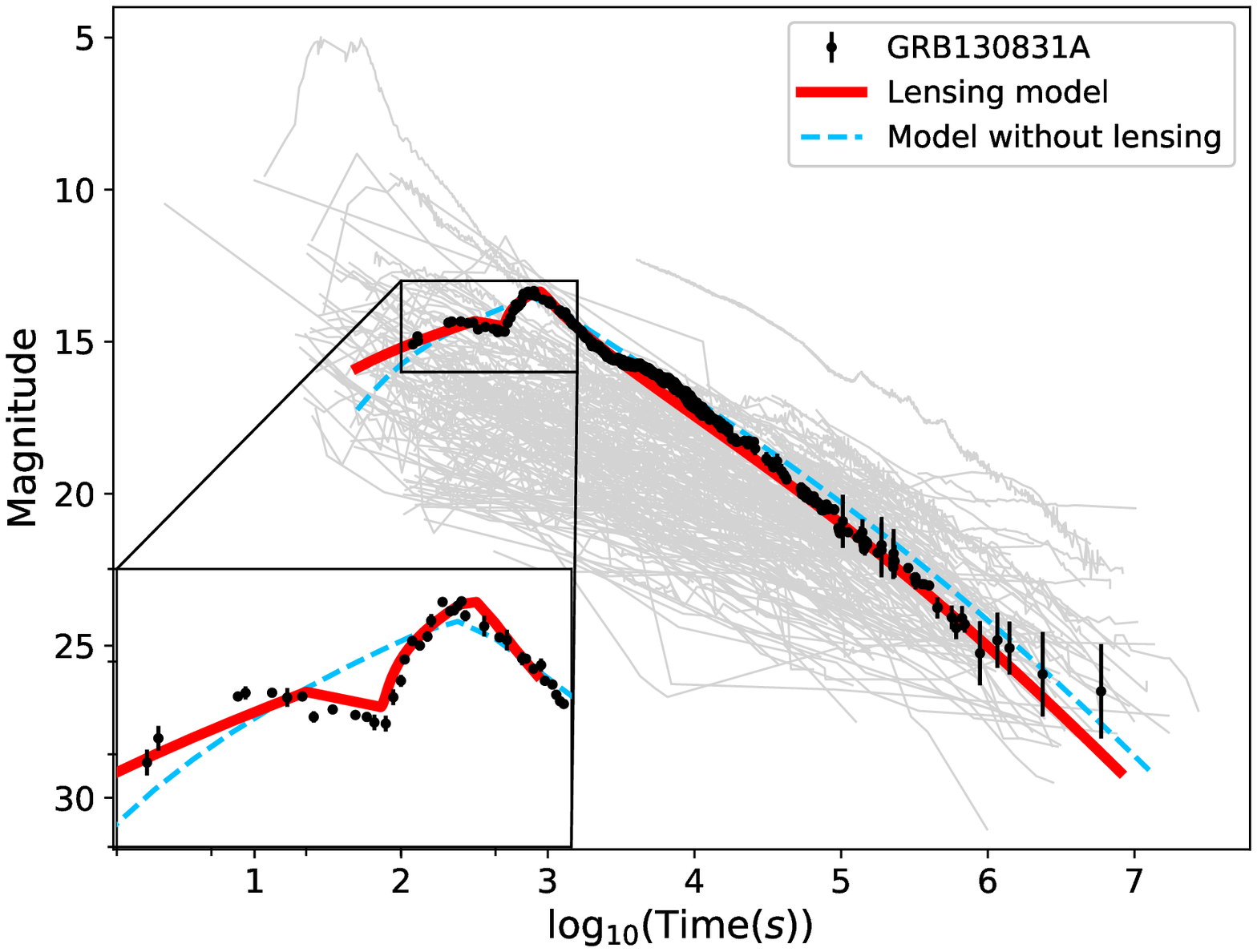}
        \includegraphics[width=8.5cm,height=7cm]{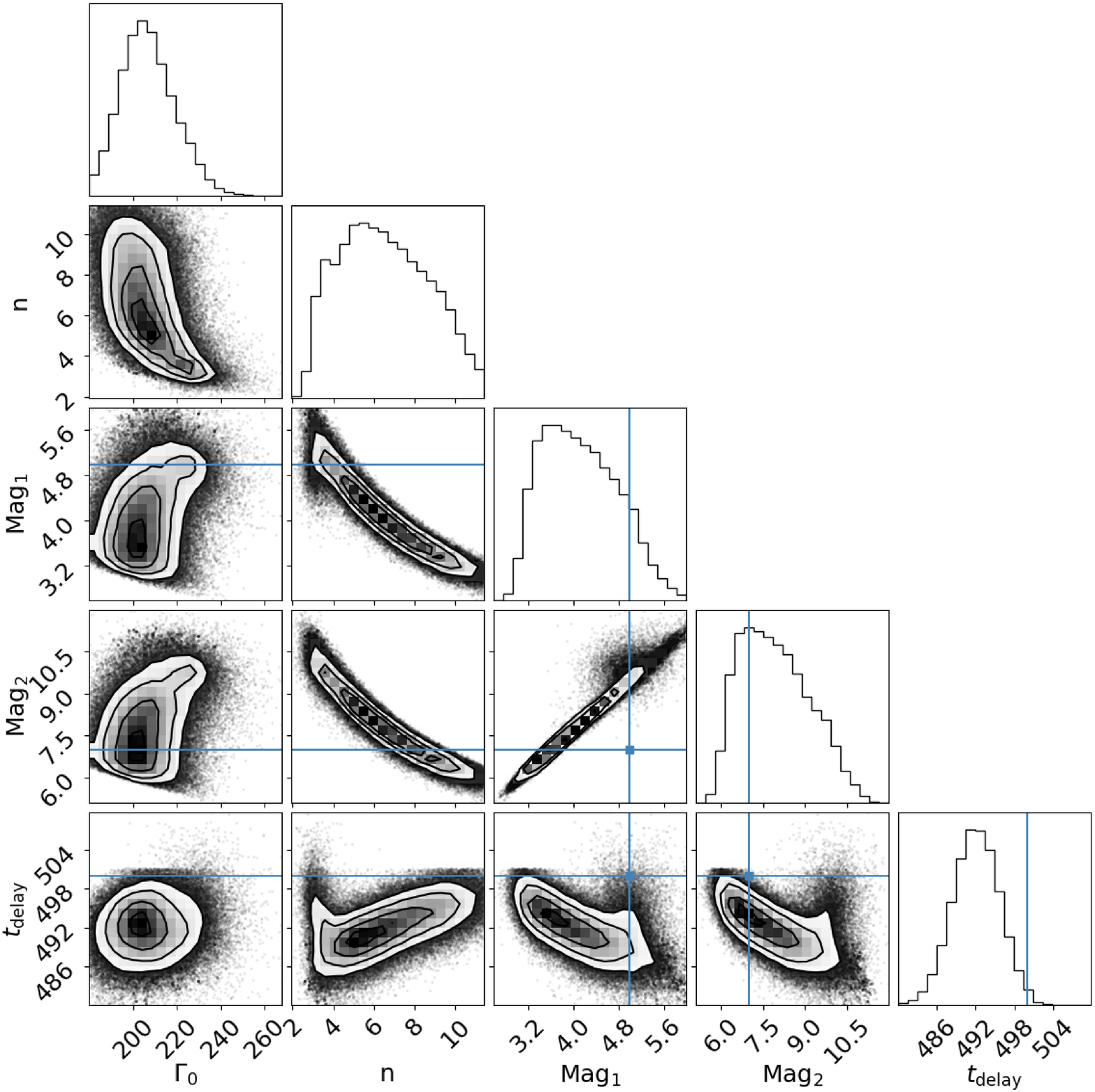}\\
        \includegraphics[width=8.0cm,height=6.5cm]{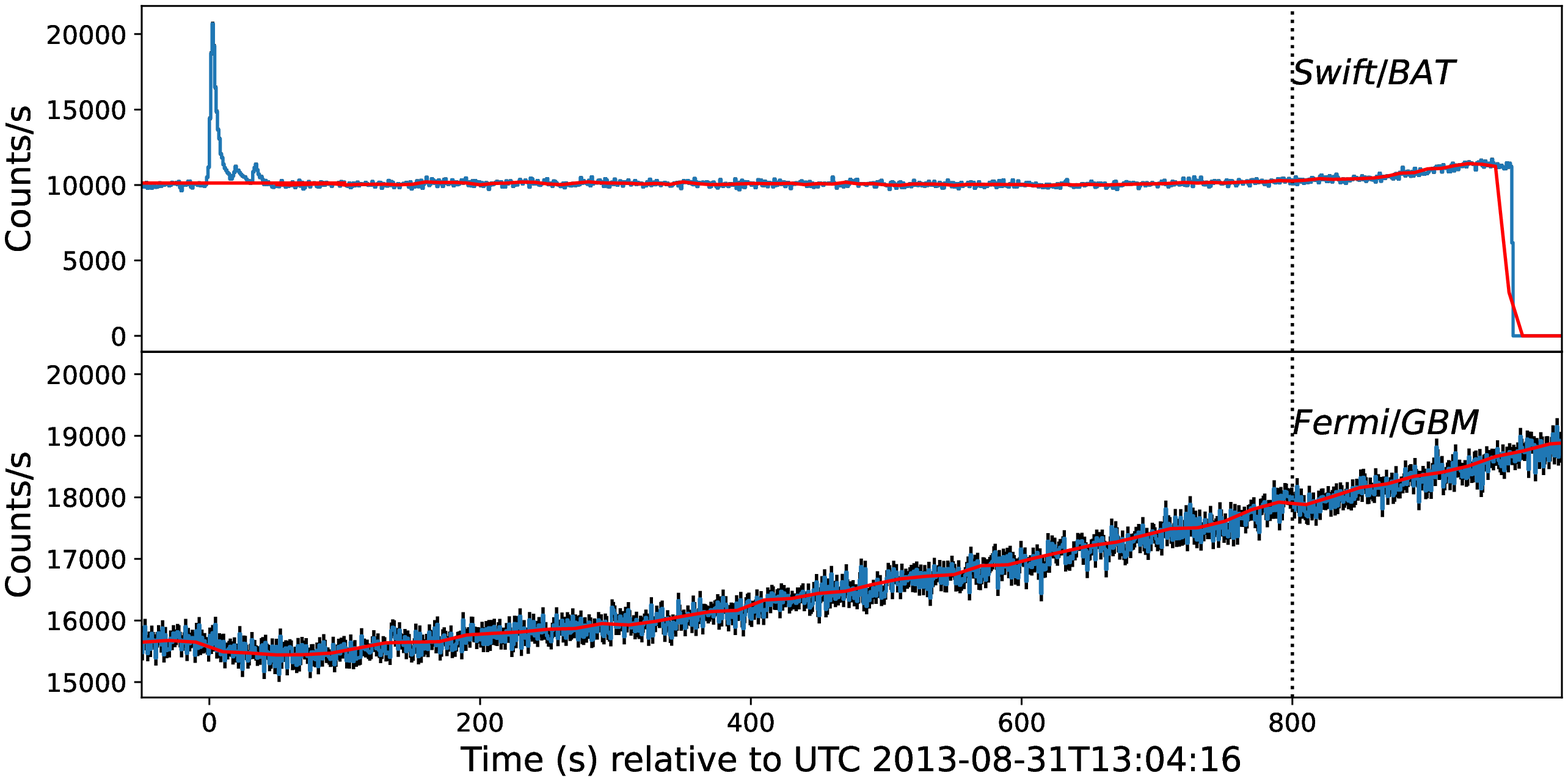}
        \includegraphics[width=8.5cm,height=7cm]{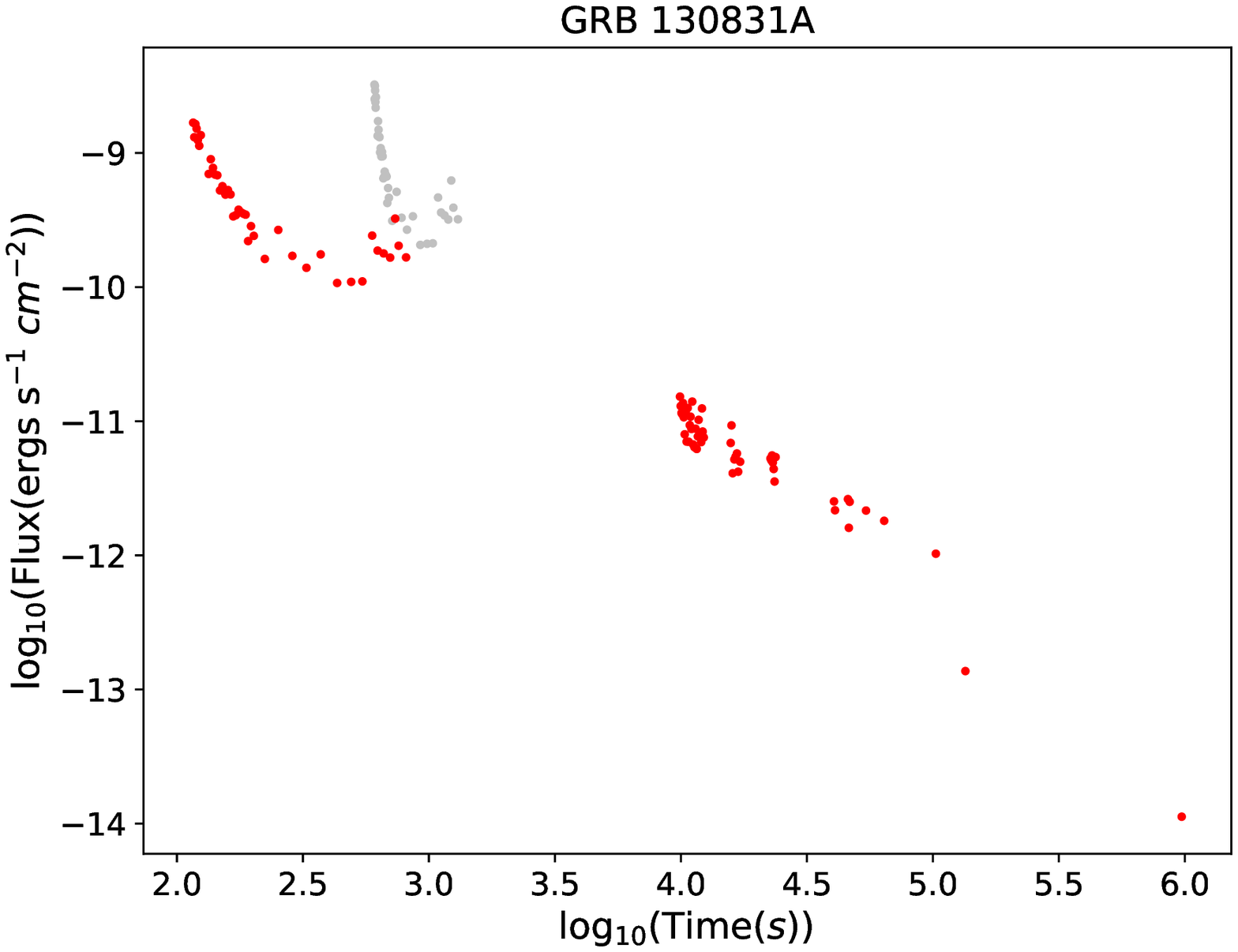}

    	\caption{Top panels: the best fitting results for the GRB 130831A optical afterglow, the corresponding corner plot of the posterior probability distribution for the fitting. The gray background lines show the optical lightcurves collected in our sample. Bottom panels: gamma-ray and X-ray band observations of GRB 130831A, where red lines in the left panel present the background level. In the plot of X-ray observation, for comparison, we manually shift the early X-ray data by $t_{\rm delay}=492$s and $\rm Mag_2/Mag_1=1.9$ to show the expected flux level around the time when the second optical peak emerging (see grey dots in the right panel). }
    	\label{fittingresult}	
\end{figure}

Based on the predicted characteristics in Section 2.3, we analyze optical band by searching for the presence of additional signal due to gravitational lensing of GRBs. We extensively search for the optical data from published papers or from the Gamma-ray Coordinates Network (GCN) Circulars if no published paper is available. We totally found 358 GRBs with optical observations being reported from 1997 February to 2020 December, including 308 GRBs having well-sampled optical light curves, which contain at least 3 data points, excluding upper limits \footnote{Part of our sample have been published in the literature \cite[e.g.][]{li12,wang15}.}. By visual inspection, we systematically searched through our sample and identify 3 bursts that show a clear onset peak, followed by a steep rebrightening. These are GRB 100621A, GRB 100901A and GRB 130831A. It is worth noting that some sources are also found to contain a steep rebrightening signature, but their early onset peak has not been observed (some examples are collected in \cite{Ugarte Postigo18}). These sources together with GRB 100621A and GRB 100901A are further excluded because their rebrightening peak is too wide to be explained by lensing effect, but more like a giant bump caused by fallback accretion process \citep{wu13, zhao21}. Overall, only one possible candidate is left: GRB 130831A with a redshift of 0.479 \citep{Cucchiara13}. \cite{De Pasquale16} provided a comprehensive analysis for the ultraviolet, optical and infrared photometry of GRB 130831A, whose combined optical light-curves show a rapid temporal evolution between 400 and 800 s after the trigger (the optical flux suddenly increased by a factor of $\sim$5). The sharp rise feature is very difficult to be explained within the external shock framework, unless invoking some complicate energy injection process, such as a continuously refreshed shock following the prompt emission phase \citep{ZhangQ16}. \cite{De Pasquale16} proposed that the sharp rising should be connected to internal dissipation processes which occurred in the outflow, when the Lorentz factor is very high and relativistic effects cause rapid variations of the observed flux. We are interested here in testing whether the sharp rise feature is a signature of lensing effect. Considering that the UV/optical/NIR data collected in \cite{De Pasquale16} have not been corrected for Galactic and host galaxy extinction, we use the Rc band data of GRB 130831A provided by \cite{Ugarte Postigo18}, which has been calibrated to the Vega system, corrected for Galactic extinction. In addition, the host-galaxy and supernova have been subtracted \footnote{Note that GRB 130831A is known to be associated with SN 2013fu \citep{cano14}. The Rc band data we used here has subtracted the supernova contribution.}. We apply the model described in Section 2 to fit the data of GRB 130831A, by performing the {$\chi^2$ objective function} minimization procedure with using Markov Chain Monte Carlo (MCMC) method through the emcee code \citep{f13}. In the fitting, we focus on the early observational data (e.g. $t<1500$ s), which essentially reflects whether the lensing effect exists. In order to reduce the number of free parameters in our fitting we fix several of them at their typical values. For instance, we set $E_{0}=10^{53}\ \rm{ergs}$, ${\theta _0} = 0.3$, ${R_0} = {10^{15}}$cm, ${\epsilon _e} = 0.1$, and ${\epsilon _B} = {10^{-4}}$. We take the initial bulk Lorentz factor ${\Gamma _0}$ of the outflow, the number density $n$ of the circumburst medium, the magnifications of the two images ($\rm Mag_1$, $\rm Mag_2$), the time delay between the two images $t_{\rm delay}$ as the free parameters. Note that by fitting the observed light curve, parameters characterizing the overall lensing effect (magnifications and time delay) can be obtained directly. The lens system parameters, such as $M_l$, $z_l$, $\beta$, could hardly be resolved due to strong degeneracy, unless some of them ($z_l$ or $\beta$) could be directly determined with observations. Figure \ref{fittingresult} shows our best fitting results for the lightcurve and the corresponding corner plot of the posterior probability distribution for the fitting.  
We can see that the early lightcurve of GRB 130831A could be well fitted with our proposed model. The best fit parameter values at the $1\sigma$ confidence level are ${\Gamma _0} = 206_{ - 11}^{ + 13}$, $n = 6.3_{ - 2.2}^{ + 2.6}\;{\rm{c}}{{\rm{m}}^{{\rm{ - 3}}}}$, ${\rm{Ma}}{{\rm{g}}_{\rm{1}}}{\rm{ = 4}}.1_{ - 0.6}^{ + 0.8}$, ${\rm{Ma}}{{\rm{g}}_{\rm{2}}}{\rm{ = 7}}{\rm{.8}}_{ - 1.2}^{ + 1.5}$, ${t_{{\rm{delay}}}} = 492_{ - 3.6}^{ + 3.5}$ s. Constraints on all parameters associated with the external shock are consistent with typical values. Note that due to the lack of information about the baseline (unlensed) flux, the values of $\rm Mag_1$ and $\rm Mag_2$ should be highly degenerated with the afterglow modeling parameters, so that the ratio between $\rm Mag_1$ and $\rm Mag_2$ is more meaningful, compared with their absolute values. According to the constraint result, the time delay between the two lensed images is $ \sim 500$ s and the flux ratio is $\sim2$. Through external interpolation, we show that the best fitting lightcurve is generally consistent with the late time observations\footnote{Between 3000 s and 17000 s, the observational data show a plateau feature, which makes this part of the data slightly higher than the extrapolation extension of the best fitting lightcurve. This is acceptable since the plateau feature is likely related to the late central engine activity (see \cite{De Pasquale16} for detailed physical interpretations for late time observations of GRB 130831A). }. For completeness, we also fit the early data of GRB 130831A by using the standard afterglow model without invoking any lensing effect, and the best fit result is shown in Figure \ref{fittingresult} as well. We then use the Bayesian method to test how significant the observation supports the lensing model (marked as model $M_1$) against the model without strong lensing (marked as model $M_2$). It turns out the lensing model is more preferable by the observations (see Appendix for details of our adopted Bayesian method and relevant results). 

However, it is worth noting that although the optical data of GRB 130831A can be well fitted by the lens effect, it is highly doubtful whether the source is actually lensed for the following reasons:
\begin{itemize}
\item The second arriving image has a larger magnification. Usually, for typical lens models, if there are only two lensed images, the second image should be fainter as the echo. In other words, even if GRB 130831A is actually lensed, it may have experienced very special gravitational lensing events. One possible lens model is a point mass+shear which might explain the brighter trailing images \citep{chen21}. Another plausible explanation for such a particular combination of time delay and the flux ratio could be the existence of dark matter subhalo of mass $\sim10^6M_\odot$ in the galaxy resulting in the obtained value of time delay. In this case the magnification of the second image (and estimated value of the flux ratio) could be caused by microlensing effect, i.e. by the stars in the lens galaxy. Let us note that microlensing effect on GRB afterglows has been discussed in some previous works \cite[e.g.][]{koopmans01,gaudi01,ioka01}.
\item Since GRB 130831A was triggered by Swift, we checked the historical data of BAT and XRT. As shown in Figure \ref{fittingresult}, when the second optical peak emerging, there is no signal showing up in the BAT data. Before the data recording stopped, the BAT data did show an increasing trend, but it is difficult to judge whether it is a real signal or it is caused by satellite attitude adjustment. There is an X-ray flare starting at about 500 s after the trigger, which is roughly consistent with the optical rising feature, but the flux at $\sim$ 800 s is about an order of magnitude lower than the flux at $\sim$ 100 s, inconsistent with the expected magnification ratio. We also checked the historical data of Fermi/GBM and found that the position of GRB 130831A was blocked by the Earth at the trigger time, but it entered the monitoring region of GBM after hundreds of seconds. As shown in Figure \ref{fittingresult}, when the second optical peak was emerging, there were no signal showing up in the GBM data. Interestingly, the GBM data also showed an obvious rise around the time when the second optical peak emerged. However, this may be caused by the rise of the background, because data in all the detectors of GBM have risen, and the satellite has entered the high latitude region in that period.
\end{itemize}

\section{Discussion and Conclusion}\label{conclusion}

It has long been proposed that GRBs have potential to be gravitationally lensed into multiple images, due to their high redshift nature. The detection of lensed GRBs could be used to improve constraints on cosmological parameters such as the Hubble constant ${{\rm{H}}_{\rm{0}}}$, to test fundamental physics from the propagation speed, and to make constraints on the abundance of compact dark matter, and so on \cite[][for a review]{oguri19}.

The traditional searches for lensed GRBs are focused on the gamma-ray band. Here we propose that one can search for lensed GRBs by combining gamma-ray and the multi-band afterglow data. We use the standard afterglow model and two standard lens models to calculate the characteristics of the lensed afterglow lightcurves in different scenarios concerning lens mass and time delays. Based on our results, we suggest that the future search for lensed GRBs can be focused on several following cases. 1) When a lens is a compact object, the time delay is in order of $\sim100$s or less. In this case, multiple gamma-ray images could be either separated or overlapped, depending on the comparison between $T_{90}$ and $t_{\rm delay}$. It is also possible that only one image signal can be seen, once the later arriving signal is under the detector threshold. In this case, we can use the observation of X-ray and optical afterglow data to judge whether these GRBs are lensed or they just happen to have several similar emission episodes. If the GRBs are indeed lensed, the X-ray afterglows are likely to contain several X-ray flares with similar width in linear scale (the later, the narrower in logarithmic scale) and similar spectrum. Moreover, the optical afterglow lightcurve will show rebrightening signatures. 2) When the compact object is a low mass galaxy, the time delay would be $\sim10^3-10^4$s. In this case, there will be two independent GRB triggers in the gamma ray band, and there will be a sharp rise flare feature in both X-ray and optical afterglow lightcurve, which can help to easily justify the lensing effect. 3)  When the compact object are massive galaxies, the time delay would be in order of days or even longer. In this case, there will be two independent GRB triggers in the gamma-ray band, and the late X-ray signal will rise sharply. Nevertheless, optical telescopes now usually have the ability to resolve different lensed images, so as to better help confirm the lensing effect. Note that for all three situations, radio telescopes usually have the ability to resolve multiple images. Therefore, for suspected samples selected from gamma-ray, X-ray, and optical bands, concerted radio observations could finally help to confirm whether they are real lensing events.

Through archive data searching in optical band, we find one potential candidate of lensed GRB, 130831A, with a time delay  $\sim$500 s, which should belong to the first situation we discuss above. However, the gamma-ray and X-ray band observations of GRB 130831A seem not to support the lensing hypothesis. Therefore, whether this source is a lensed event remains to be further discussed. With the successful operation of many sky survey projects in multiple bands, especially the establishment of all-sky gamma-ray monitors, more lensed GRBs would be detected and accurately certified in the future.

\acknowledgments
We thank Dr. David Alexander Kann for sharing with us the data of GRB 130831A. We thank the anonymous referee for the helpful comments that have helped us to improve the presentation of the paper. This work is supported by the National Natural Science Foundation of China (NSFC) under Grant No. 11690024, 12021003, 11633001, 11973034.

\appendix

Here we employ the {\tt{dynesty}} nested sampling algorithm \citep{Skilling12} implemented in {\tt{bilby}} \citep{ashton19} to calculate the Bayes factor 
\begin{equation} \label{BF}
{\rm BF}^{1}_{2}=\frac{{{\cal{Z}}_1}({d|M_{1})}}{{{\cal{Z}}_2}( d|M_{2})}\, ,
\end{equation}
where ${\cal{Z}}_{1,2}$ is the evidence of model $M_{1,2}$, which can be estimated as 
\begin{equation}
{\cal{Z}}({d|M})=\int {\cal{L}}(d|\theta,M)\pi (\theta|M)d\theta\, ,
\end{equation}
where $\cal{L}$ is the likelihood and $\pi$ is the prior of parameters. Here we choose Gaussian likelihood and three different prior cases (P1, P2 and P3) for both $M_1$ and $M_2$. The results are shown in Table.{\ref{tab:bf}}. It turns out $\ln({\rm{BF}^{1}_{2}})$ between lensing and non-lensing model could be as large as $\sim$600, and this result is not sensitive to the choice of prior. Normally when $\ln({\rm{BF}^{1}_{2}})$ is larger than eight, one can conclude that $M_1$ model is significantly favoured against $M_2$ model \citep{mackay03}. In our case, it is clear that the lensing model is more preferable by the optical observations. In order to check the results from {\tt{dynesty}}, we also calculated $\rm{ln(BF^1_2)}$ by using other public codes, such as {\tt{ptemcee}} \citep{Vousden16,f13}, whose results are consistent with the result from {\tt{dynesty}}. The seemingly excessive large number of $\ln({\rm{BF}^{1}_{2}})$ should come from two reasons: 1) for the early data, the difference between the two models is significant, which can be clearly seen from the subgraph inserted in Figure \ref{fittingresult}; 2) the Bayes factor crucially depends on the error of observation data points (see detailed illustrations and examples in \cite{John2005}), so that the relatively small error for early data points could lead to the huge $\ln({\rm{BF}^{1}_{2}})$ value.

\begin{table}
  \centering
   \caption{The prior probability distributions adopted in the Bayesian analysis, and the natural logarithm of Bayes factor calculated by {\tt{dynesty}} for $M_1$ and $M_2$, where $\mathcal U$ and $\mathcal G$ refers to the Uniform and Gaussian type of prior.} 
    \begin{tabular}{l|c|c|cc}
      \hline\hline
          & \multicolumn{1}{c|}{P1} & \multicolumn{1}{c|}{P2} & \multicolumn{1}{c}{P3} &  \\
\hline   $\Gamma_0$ & \multicolumn{1}{l|}{$\mathcal U$(100,300)} & \multicolumn{1}{c|}{$\mathcal U$(100,500)} & \multicolumn{1}{c}{$\mathcal G
$(200,40)} &  \\
    $n$     & \multicolumn{1}{c|}{$\mathcal U$(0,5)} & \multicolumn{1}{c|}{$\mathcal U$(0,50)} & \multicolumn{1}{c}{$\mathcal G$(10,5)} &  \\
    $\rm Mag_1$ & \multicolumn{1}{c|}{$\mathcal U$(0,6)} & \multicolumn{1}{c|}{$\mathcal U$(0,6)} & \multicolumn{1}{c}{$\mathcal U$(0,6)} &  \\
    $\rm Mag_2$ & \multicolumn{1}{c|}{$\mathcal U$(0,12)} & \multicolumn{1}{c|}{$\mathcal U$(0,12)} & \multicolumn{1}{c}{$\mathcal U$(0,12)} &  \\
    $\rm t_{delay}$ & \multicolumn{1}{c|}{$\mathcal U$(450,550)} & \multicolumn{1}{c|}{$\mathcal U$(450,550)} & \multicolumn{1}{c}{$\mathcal U$(450,550)} &  \\
\hline $\rm{ln(BF^1_2)}$
& 589   & 596   & 597   &  \\
\hline\hline
    \end{tabular}%
  \label{tab:bf}%
\end{table}%

\end{document}